\documentclass[aps,showpacs,twocolumn,superscriptaddress]{revtex4}

\usepackage{amsmath}
\usepackage{amssymb}                 
\usepackage{wasysym}
\usepackage[dvips]{graphicx}
\DeclareGraphicsExtensions{.eps}

\newcommand{\ri}{{ \rm i }}
\newcommand{\re}{{ \rm e }}
\newcommand{\rd}{{ \rm d }}
\newcommand{\rmi}{{ \rm i }}

\newcommand{\E}{{ \mathcal{E} }}
\newcommand{\be}{\begin{equation}}
\newcommand{\ee}{\end{equation}}
\newcommand{\nn}{\nonumber}

\begin{document}

\title{Bose-Einstein condensates in accelerated double-periodic optical lattices:\\
Coupling and Crossing of resonances}
\author{D. Witthaut}
\email{witthaut@physik.uni-kl.de}
\affiliation{Fachbereich Physik, TU Kaiserslautern, D--67653 Kaiserslautern, Germany}
\author{E. M. Graefe}
\affiliation{Fachbereich Physik, TU Kaiserslautern, D--67653 Kaiserslautern, Germany}
\author{S. Wimberger}
\affiliation{Dipartimento di Fisica Enrico Fermi and CNR-INFM,\\
Universit\'a degli Studi di Pisa, Largo Pontecorvo 3, I-56127 Pisa, Italy}
\affiliation{CNISM, Dipartimento di Fisica del Politecnico,\\
Corso Duca degli Abbruzzi 24, 10229 Torino, Italy}
\author{H. J. Korsch}
\affiliation{Fachbereich Physik, TU Kaiserslautern, D--67653 Kaiserslautern, Germany}
\date{\today}

\begin{abstract}
We study the properties of coupled linear and nonlinear resonances.
The fundamental phenomena and the level crossing scenarios are introduced
for a nonlinear two-level system with one decaying state, describing
the dynamics of a Bose-Einstein condensate in a mean-field
approximation (Gross-Pitaevskii or nonlinear Schr\"odinger equation).
An important application of the discussed concepts is the dynamics of
a condensate in tilted optical lattices. In particular the properties of
resonance eigenstates in double-periodic lattices are discussed, in the
linear case as well as within mean-field theory. The decay is
strongly altered, if an additional period-doubled lattice is introduced.
Our analytic study is supported by numerical computations of nonlinear
resonance states, and future applications of our findings for experiments
with ultracold atoms are discussed.
\end{abstract}

\maketitle


\section{Introduction}
\label{sec-intro}

In the last decade, the advance of atom and quantum optics has made it possible
to realize and to study the evolution of the center-of-mass motion on scales
ranging from the microscopic (single particle) to the macroscopic
(many-particle) realm \cite{Adv06,MO2006}.
In a typical experiment with ultracold
atoms, interactions can either be made negligibly small or reduced to
a mean-field effect on the evolution of the macroscopic order parameter of
a Bose-Einstein condensate (see, e.g., \cite{MO2006,Pita03} and
references therein). The latter approach results in an effective
nonlinear Schr\"odinger equation, the following
Gross-Pitaevskii equation:
\be
  \left( - \frac{\hbar^2}{2m} \frac{\partial^2}{\partial x^2}
  + V(x) + g |\psi(x,t)|^2 \right)
  \psi(x,t) = \ri \hbar \frac{\partial \psi(x,t)}{\partial t} \, ,
  \label{eqn-NLSE-timedep}
\ee
that describes the dynamics of the macroscopic wave function (or of the
order parameter) of a Bose-Einstein condensate (BEC) for zero temperature 
\cite{Pita03}. This mean-field description has proved to be extremely
successful and reliable for most of recent experiments.
The nonlinearity of the equation leads to a variety of surprising
phenomena, which are present even in a simple nonlinear two-level
system. Self-trapping of a BEC in a double-well trap was observed
experimentally only recently \cite{Albi05}.
The self-trapping transition manifests itself in the appearance of novel
nonlinear eigenstates \cite{Smer97}.
The appearance and disappearance of nonlinear eigenstates may also lead
to a breakdown of adiabaticity and nonlinear Zener tunneling
\cite{Wu00,Liu03,06zener_bec}.

In the present paper, we investigate nonlinear quantum dynamics in
decaying systems. Up to now, only relatively few papers have studied
nonlinear and non-hermitian quantum dynamics, discussing self-stabilizing,
shifting and broadening of nonlinear resonances
\cite{Mois04a,Schl04,04nls_delta,Wimb05,Wimb06}.
Here we focus on the coupling of nonlinear resonances in nonlinear,
non-hermitian level crossing scenarios. Our first object of investigation,
the nonlinear two-level system with one decaying level, offers analytic 
access to this subject. The eigenvalues and
eigenvalues of its linear counterpart show some interesting features,
such as exceptional crossing scenarios \cite{03crossing}.

A very natural experimental setup leading to nonlinear dynamics
and decay is the dynamics of a Bose-Einstein condensate in a tilted
or accelerated optical lattice, corresponding to the 
Wannier-Stark scenario of solid-state physics \cite{N1991}.

The decay dynamics in a nonlinear Wannier-Stark system was recently
discussed in \cite{Wimb05,Wimb06}. It was shown that a nonlinear
mean-field interaction can destroy resonant tunneling.
In this paper, we extend these studies to a double-periodic
optical lattice. The decay dynamics in this system shows some
interesting features even in the linear case, such as a splitting
of resonant tunneling peaks. The different types of non-hermitian
crossing scenarios can be observed in dependence of the system
parameters. 

The paper is organized as follows: first of all we review some
important results about the crossing scenarios in the non-hermitian
two-level system in the linear (section II) and the nonlinear (section III)
case. The double-periodic Wannier-Stark system is introduced and analyzed
in section \ref{sec-ws-lin}. Nonlinear Wannier-Stark resonances for a
doubly periodic lattice are presented in section \ref{sec-ws-nonl}.
A discussion of novel experimental applications of our findings follows 
in section VI.

\section{Crossing scenarios of resonances in linear quantum mechanics}
\label{sec-l2-lin}

We prepare for the full discussion of nonlinear resonance states as solutions
of Eq.~(\ref{eqn-NLSE-timedep}) by reviewing some essential properties
of the simpler linear case. First of all, we want to illustrate the
different types of possible curve crossing scenarios for non-hermitian
systems.
To start with, we briefly review a simple and instructive model system,
a two-level Hamiltonian with one decaying level \cite{Avro82,03crossing}:
\be
  H_2 = \left(\begin{array}{c c}
  +\epsilon - 2 \ri \gamma & v \\
  v & -\epsilon \\
  \end{array}\right)
  \label{eqn-2level-lin-ham}
\ee
with $\epsilon,v,\gamma \in \mathbb{R}$ and $\gamma \ge 0$.
In this approach it is assumed that one of the bare states decays
with rate $\gamma$, while the decay is negligible for the other one.
The mean energy of the two bare states is set to zero, the energy
difference is given by $2\epsilon$. The two states are coupled with
strength $v$.
A different, non-hermitian two-level Hamiltonian was previously discussed
by Berry \cite{Berr84a}.

\begin{figure}[t]
\centering
\includegraphics[width=8cm,  angle=0]{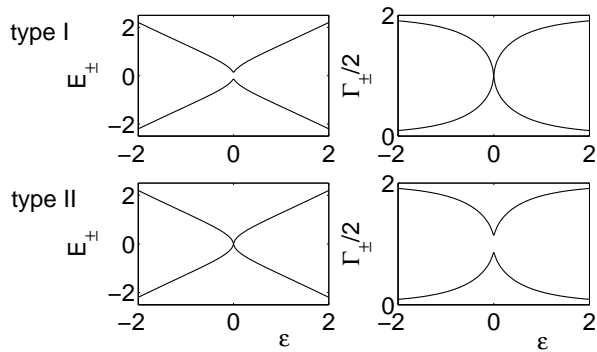}
\caption{\label{fig-2level-lin-cross}
Real (left) and imaginary (right) part of the eigenvalues (\ref{eqn-2level-lin-eig})
as a function of $\epsilon$ for $\gamma =1$. A type I crossing is found for
$v = 1.01 > \gamma$ (upper figures), a type II crossing is found for
$v = 0.99 < \gamma$ (lower figures).}
\end{figure}

The eigenvalues of the non-hermitian Hamiltonian (\ref{eqn-2level-lin-ham})
are given by
\be
  \E_\pm = -\ri \gamma \pm \sqrt{(\epsilon-\ri \gamma)^2 + v^2} = E_\pm - \ri  \Gamma_\pm/2.
  \label{eqn-2level-lin-eig}
\ee
Both real and imaginary part of the eigenvalues are different
for $\epsilon \ne 0$. (Anti) crossings of the real and imaginary part are
found only in the critical plane $\epsilon=0$. The exceptional line
$v = \pm \gamma$ separates the critical plane into different
regions:
\begin{enumerate}
\item For $|v| > \gamma$ the imaginary parts of the eigenvalues coincide,
$\Gamma_+ = \Gamma_- = 2 \gamma$, while the real parts differ.
This case is denoted as type I crossing.
\item For $|v| < \gamma$ the real parts of the eigenvalues coincide,
$E_+ = E_- = 0$, while the imaginary parts differ.
This case is denoted as type II crossing.
\item The eigenvalues are fully degenerate, $\E_+ = \E_-$, along the critical
lines $v = \pm \gamma$.
\end{enumerate}
The two different crossing types are illustrated in
Fig.~\ref{fig-2level-lin-cross}.
For a type I crossing, i.e. $|v| > \gamma$, the imaginary parts
of the eigenvalues cross while the real parts anti-cross.
For a type II crossing, i.e. $|v| < \gamma$, it is the other
way round. Physically this crossing describes a resonantly enhanced
tunneling (RET) effect: the decay rate of the lower state increases
significantly if this state is energetically close or equal to the decaying
upper level.

\begin{figure}[t]
\centering
\includegraphics[width=8cm,  angle=0]{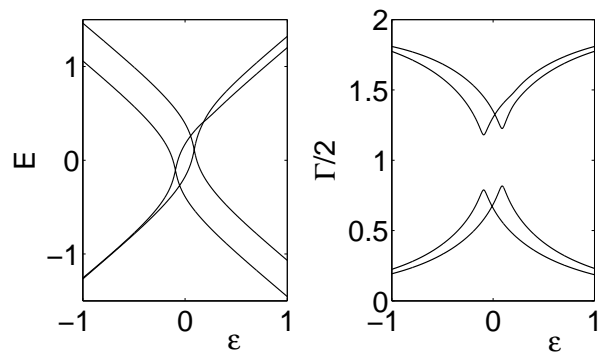}
\caption{\label{fig-4level-lin-cross1}
Real (left) and imaginary (right) part of the eigenvalues of the
four-level Hamiltonian (\ref{eqn-4level-lin-ham}) as a function
of $\epsilon$ for $\gamma =1$, $v = 0.98$, $w=0.04$ and $\delta = 0.2$.}
\end{figure}

In view of the discussion of Wannier-Stark resonances in period-doubled
lattices in the following sections we want to introduce another model
system.
We assume that the bare states split up into two states,
where the energies of the stable bare states differ slightly
by $2\delta$.
Each stable state mainly couples to one of the decaying states,
while all other couplings are assumed to be weak. We consider the Hamiltonian
\be
  H_4 = \left(\begin{array}{c c}
  H_2 + A & W \\
  W & H_2 - A\;, \\
  \end{array} \right)
  \label{eqn-4level-lin-ham}
\ee
with the two-level Hamiltonian $H_2$ defined in (\ref{eqn-2level-lin-ham}) and
\be
  A = \delta \left(\begin{array}{c c}
  0 & 0 \\ 0 & 1 \\
  \end{array} \right), \quad
  W = w \left(\begin{array}{c c}
  1 & 1 \\ 1 & 1 \\
  \end{array} \right).
\ee
Figure \ref{fig-4level-lin-cross1} shows the eigenvalues
of this Hamiltonian in dependence of the on-site energies $\epsilon$.
One observes that the resonance peak splits up into two peaks. Two
possibilities for resonant tunneling, i.e. two type II crossings, are
found instead of just one.
This crossing scenario is robust against small variations of the
coupling $w$ as long as $w \ll v$ is fulfilled. A non-vanishing
coupling $w$ causes as slight asymmetry of the two crossings.
For $v > \gamma$ one has two type I crossings instead, i.e., the
imaginary parts of the eigenvalues cross while the real parts
anti-cross.

\begin{figure}[t]
\centering
\includegraphics[width=8cm,  angle=0]{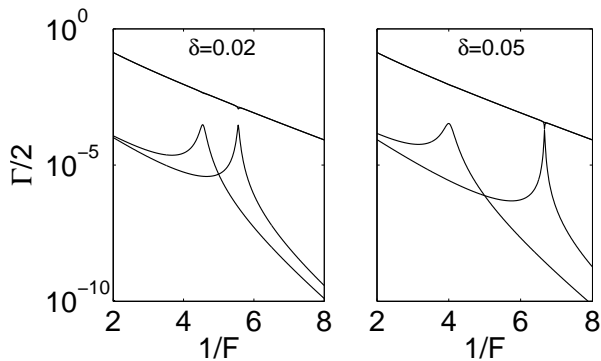}
\caption{\label{fig-4level-lin-cross2}
Imaginary part of the eigenvalues of the four-level Hamiltonian
(\ref{eqn-4level-lin-ham}) as a function of $1/F$ and
$\delta = 0.02$ (left) and $\delta = 0.05$ (right).
See text for details.}
\end{figure}

The change of a system parameter, e.g. the strength of the Stark
field $F$ in the Wannier-Stark system discussed in section \ref{sec-ws-lin},
will typically affect the bare state energies $\epsilon$ as well as the
decay rate $\gamma$ and the coupling strengths.
Therefore we consider a variety of the four-level Hamiltonian
(\ref{eqn-4level-lin-ham}), the parameters of which are functions
of the external field $F$:
\begin{eqnarray}
  \epsilon &=& -F/2 + 0.1 \nn \\
  \gamma &=& F \, \re^{-1/F} \nn \\
  v &=& 0.05 \, F \, \re^{-1/2F} \nn \\
  w &=& 0.01 \, F \, \re^{-1/2F}\;.
  \label{eqn-wsnlevel-para}
\end{eqnarray}
The exponential scaling of the decay rate $\gamma$ is well known from
standard Landau-Zener theory (see, e.g., \cite{Holt00}).
The dependence of the bare state energies and the coupling coefficients
on $F$ were analyzed in detail for a two-ladder system in \cite{00restun}.
It was shown that the Wannier-Stark spectrum is accurately described
assuming that the parameters scale as in Eq.~(\ref{eqn-wsnlevel-para}).
The actual values of coefficients in (\ref{eqn-wsnlevel-para})
are chosen in an ad hoc manner for illustration only. The resulting
decay rates are illustrated in Fig.~\ref{fig-4level-lin-cross2}.
For $\delta = 0.02$, we find two type II crossings. With increasing
$\delta$, the crossing on the right changes its form and becomes a type I
crossing. Such a crossing scenario is naturally realized for Wannier-Stark
resonances in double-periodic lattices, as will be shown in
Fig.~\ref{fig-wsres-crossing} below.

\section{Nonlinear non-hermitian crossing scenarios}
\label{sec-l2-nonl}

The linear non-hermitian two-level system described in section~\ref{sec-l2-lin} neglects any particle-interaction. Including this interaction in a meanfield-description according to the Gross-Pitaevskii-equation~(\ref{eqn-NLSE-timedep}) yields a nonlinear non-hermitian two-level system \cite{06nlnh}, described by the Hamiltonian
\be
H_{\rm mf}=\left(\begin{array}{c c}
   \epsilon+ 2c |\psi_1|^2 -2 \rmi \gamma & v \\
   v & -\epsilon + 2c |\psi_2|^2  \\
   \end{array} \right),
   \label{eqn-2level-nlin-ham}
\ee
where the nonlinearity parameter $c$ is proportional to the parameter $g$ in the Gross-Pitaevskii-equation~(\ref{eqn-NLSE-timedep}).
The nonlinear eigenstates are then defined as the self-consistent
solutions of the time-independent Gross-Pitaevskii-equation
\be
  H_{\rm mf}\left( \begin{array}{c} \psi_1 \\ \psi_2 \\ \end{array} \right)
   = \mu \left( \begin{array}{c} \psi_1 \\ \psi_2 \\ \end{array} \right).
   \label{eqn-ev2-nlin-as}
\ee
The nonlinear eigenstates crucially depend on the normalization of the
state vector, which is fixed as $|\psi_1|^2+|\psi_2|^2=1$ throughout this
section.
For convenience we symmetrize the nonlinear Hamiltonian~(\ref{eqn-2level-nlin-ham})
by substracting a constant energy term $c(|\psi_1|^2 + |\psi_2|^2)$.
The Gross-Pitaevskii-equation (\ref{eqn-ev2-nlin-as}) then reads
\be
\left(\begin{array}{c c}
   \epsilon+c\kappa -2 \rmi \gamma & v \\
   v & -\epsilon -c \kappa \\
   \end{array} \right)
   \left( \begin{array}{c} \psi_1 \\ \psi_2 \\ \end{array} \right)
   = \mu \left( \begin{array}{c} \psi_1 \\ \psi_2 \\ \end{array} \right)
   \label{eqn-ev2-nlin}
\ee
with $\kappa=|\psi_1|^2-|\psi_2|^2$.
The self-consistent solutions of this nonlinear equation define the
nonlinear eigenstates and eigenvalues.

\begin{figure}[t]
\centering
\includegraphics[width=8cm,  angle=0]{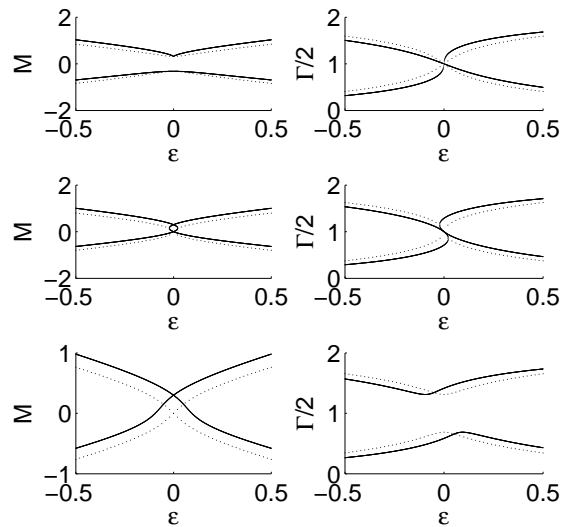}
\caption{\label{fig-2level-nlin-cross12weakc}
Real (left) and imaginary (right) part of the eigenvalues (\ref{eqn-ev2-nlin})
as a function of $\epsilon$ for $\gamma =1.00$ and $c=0.3$ and $v=1.05 >\gamma$
(upper figures), $v=1.00=\gamma$ (middle figures), $v=0.95<\gamma$ (lower figures).
The linear levels ($c=0$) are plotted as dotted lines for comparison.}
\end{figure}

Note that the nonlinear eigenstates are not connected to stationary solutions
of the time dependent system, if the chemical potential turns out to be complex,
since the dynamics depends crucially on the normalization of the state-vector,
which is not constant for a complex valued chemical potential.
After some algebraic manipulation one can show that the nonlinear eigenstates,
i.e., the solutions of equation (\ref{eqn-ev2-nlin}) are given by the real roots
of the equation
\be
  (c^2+\gamma^2)\kappa^4 + 2c \epsilon \kappa^3 + (v^2+\epsilon^2-\gamma^2-c^2)\kappa^2
  -2c\epsilon \kappa-\epsilon^2=0.
  \label{eqn-nlin-nherm-ew-pol}
\ee
Depending on the parameters, there are two or four real roots and each
of them is connected to a complex eigenvalue by
\be
  \mu=c+\epsilon/\kappa-\rmi \gamma (1+\kappa) = M-\rmi \Gamma/2.
\ee
For $\epsilon \to \pm \infty$, the linear term dominates and one has only two
eigenvalues. For $\epsilon=0$, $\kappa=0$ is a double degenerate solution
of equation (\ref{eqn-nlin-nherm-ew-pol}).
For $\gamma < v$ these states are connected to the common linear
(anti-)symmetric eigenstates, while this is not the case for
$\gamma>v$. In the following we consider the crossing
scenario of the eigenvalues in dependence on $\epsilon$ for different fixed
values of the other parameters.

The nonlinear eigenstates of a two-level system are well known for
the hermitian case $\gamma = 0$ \cite{Wu00,Liu02,Liu03}.
Novel eigenstates emerge with broken symmetry if the nonlinearity
exceeds a critical value, $|c| > c_{\rm cr} = v$,
which is given by the coupling strength $v$ which corresponds to half of the gap
between the linear levels at $\epsilon=0$. The levels show looped structures
around $\epsilon=0$ with a width $|\epsilon| \le (c^{2/3} - v^{2/3})^{3/2}$.

\begin{figure}[t]
\centering
\includegraphics[width=8cm,  angle=0]{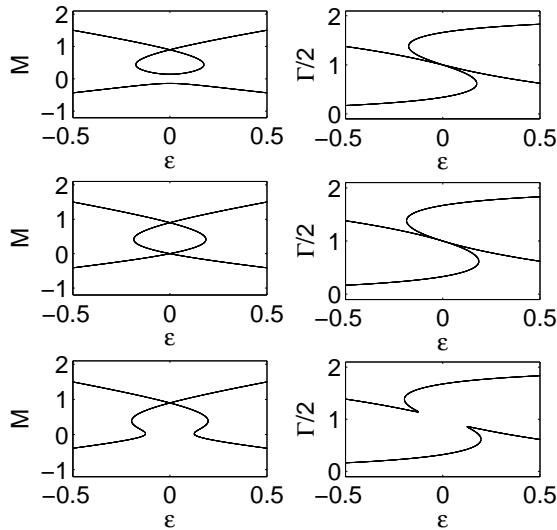}
\caption{\label{fig-2level-nlin-cross12strongc}
Real (left) and imaginary (right) part of the eigenvalues (\ref{eqn-ev2-nlin})
as a function of $\epsilon$ for $\gamma =1.00$ and $c=0.9$ and $v=1.01 >\gamma$
(upper figures), $v=1.00=\gamma$ (middle figures), $v=0.99<\gamma$ (lower figures).}
\end{figure}

Let us first discuss the effect of a weak nonlinearity:
Figure~\ref{fig-2level-nlin-cross12weakc} shows the eigenvalues
for a relatively weak nonlinearity in comparison with the linear
case $c=0$. The size of the gaps is not altered by the nonlinearity,
which leads to the important fact that the nonlinearity does {\it not}
influence the crossing type. Nevertheless it changes the shape of
the levels. For a type I crossing, the real part of the upper level
is sharpened while the one of the lower level is flattened,
which is well-known for $\gamma = 0$ \cite{Wu00}. For a type II crossing the
effect is basically the same, but is accompanied by a shift of
the crossing point from $\mu=0$ to $\mu = c$ and the lower level
is stretched to this point. The imaginary parts bend slightly to
the left.
At the exceptional point additional eigenvalues emerge in a narrow
interval around $\epsilon=0$.

In general, the presence of moderate decay facilitates the formation of novel
eigenstates. Figure \ref{fig-2level-nlin-cross12strongc} shows the
nonlinear levels for $c = 0.9$, which is slightly below the critical
value $c_{\rm cr} = v$ for $\gamma=0$.
For a type I crossing, $v > \gamma$, the real parts show a familiar
loop. The imaginary parts cross as usual, showing an additional
S-shaped structure. In fact, the critical nonlinearity for the emergence
of looped levels is decreased to
\be
  c_{\rm cr}=\sqrt{v^2-\gamma^2},
  \label{eqn-nlin-nherm-c-crit}
\ee
which can be seen by analyzing the behavior of the polynomial
(\ref{eqn-nlin-nherm-ew-pol}) for $\epsilon=0$. At the exceptional
point $v=\gamma$, the critical nonlinearity tends to zero and there are
additional eigenstates even in the case of arbitrary weak nonlinearity.
For a type II crossing, $v <\gamma$, the imaginary parts anti-cross
and the real parts cross in a manner which can be understood as a
destruction of the loop at its lower edge. The crossing appears
at the former crossing point of the loop at $\epsilon=0$ and
$\mu=c\neq0$. In the non-decaying case, $\gamma=0$, novel eigenstates
firstly emerge at the point  of the avoided crossing at $\epsilon=0$.
This remains true for a type I crossing, $v > \gamma$. For a type II
crossing, $v < \gamma$, however, novel eigenvalues emerge, again in an
S-shaped structure around some non-zero value of $\epsilon$.

Concluding this section, a weak nonlinearity does not alter the crossing type,
however it deforms the levels in a characteristic manner. For a type II crossing
the real parts cross at $\mu=c\neq 0$. At the exceptional point novel eigenstates
emerge, even for small nonlinearities. The presence of decay facilitates the
formation of novel eigenstates for stronger nonlinearities. For type I crossings,
loops appear if $|c| > c_{\rm cr} = \sqrt{v^2-\gamma^2}$. For a type II crossing
the additional eigenstates emerge around some nonzero value of $\epsilon$ forming
a double S-structure. If the sign of the nonlinearity is changed, the levels
interchange their behavior, i.e., the real parts are mirrored at the $\epsilon$-axis,
the imaginary parts at the $\Gamma$-axis.

\section{Wannier-Stark resonances in double-periodic lattices}
\label{sec-ws-lin}

\subsection{Fundamentals of the linear Wannier-Stark system}

A prime example for resonances and resonant tunneling is the
(linear) Wannier-Stark problem described by the Hamiltonian
\be
    H_{WS} = -\frac{1}{2} \frac{\partial^2}{\partial x^2}
    +V(x) + F x
    \label{eqn-ws-ham}
\ee
with a periodic potential $V(x+d) = V(x)$. We use rescaled units in
which $\hbar = M = 1$.
The Wannier-Stark problem was already discussed in the early days of
quantum mechanics in the context of electrons in solids under the
influence of an external electric field \cite{Bloc28}.
Coherent dynamics of electrons in semiconductor superlattices were
observed not until the 1990's. Experiments showed Bloch oscillations
for 'weak' electric fields and decay for stronger fields
\cite{Feld92,Leo03}. The Wannier-Stark system is furthermore
realized for the propagation of light pulses
in thermo-optically biased coupled waveguides. 
Bloch oscillations as well as decay could thus be observed 
directly in real space \cite{Pert99}. On the other hand,
recent experiments with cold atoms and Bose-Einstein condensates in
optical lattices offer some considerable advantages 
\cite{Daha96,MO2006,Pisa_PRL01,Ande98}.
Scattering by lattice defects or impurities is absent and experimental
parameters can be tuned in a wide range. The periodic potential is
generated by a standing laser beam and thus simply cosine shaped.

\begin{figure}[tb]
\centering
\includegraphics[width=8cm,  angle=0]{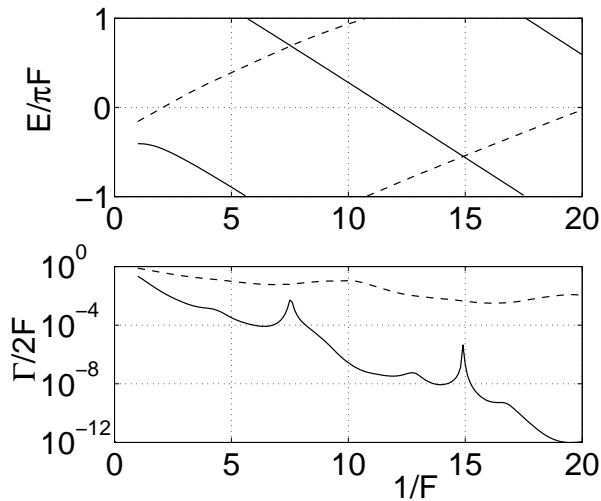}
\caption{\label{fig-restunnel0}
Eigenenergies (upper panel) and decay rates (lower panel) of the
two most stable Wannier-Stark ladders for the
potential $V(x) = \cos x$.}
\end{figure}

Let us briefly review some fundamentals of Wannier-Stark resonances,
which are defined by the eigenvalue equation
\be
   H_{WS} \Psi_{\alpha,n}(x) = \E_{\alpha,n} \Psi_{\alpha,n}(x).
\ee
Here, $\alpha$ is the ladder index and $n \in \mathbb{Z}$ labels
the lattice sites. The Wannier-Stark Hamiltonian (\ref{eqn-ws-ham})
is non-hermitian due to the boundary condition: A wave packet will
eventually decay towards $x \rightarrow -\infty$. In fact, it has been
shown that the spectrum of the Hamiltonian is continuous with
embedded resonances \cite{ZakWannAvro}. Thus the
resonance eigenenergies are complex,
$\E_{\alpha,n} = E_{\alpha,n} - \ri \Gamma_{\alpha}/2 $,
where the imaginary part $\Gamma$ gives the decay rate.
The Wannier-Stark Hamiltonian has one important symmetry, it is
invariant under a simultaneous spatial translation over a lattice
period $d$ and an energy shift $d F$. This symmetry is expressed
by the commutation relation
\be
  [H_{WS},T_m] = -m dF T_m,
\ee
where $T_m$ is the translation operator over $m$ lattice periods.
Now it is easy to see that the Wannier-Stark resonances from one
ladder $\alpha$ are related by a simple translation,
\begin{eqnarray}
  && H_{WS} T_m \Psi_{\alpha,n}(x) \nn \\
  && \quad = T_m H_{WS} \Psi_{\alpha,n}(x) + [H_{WS},T_m] \Psi_{\alpha,n}(x) \nn \\
  && \quad = \left(\E_{\alpha,n} - m dF\right) T_m \Psi_{\alpha,n}(x).
\end{eqnarray}
Thus the discrete spectrum is arranged in the form of the so-called
Wannier-Stark ladders,
\begin{eqnarray}
  && \Psi_{\alpha,n}(x) = \Psi_{\alpha,0}(x-nd) \nn \\
  && \E_{\alpha,n} = E_{\alpha,0} + ndF - \ri \Gamma_{\alpha}/2.
\end{eqnarray}
The different ladders are labelled by $\alpha = 0,1,2,\hdots$.
An efficient method to calculate the resonance eigenstates was introduced
in \cite{99trunc}, a recent review can be found in \cite{02wsrep}.

The decay rate $\Gamma_{\alpha}$ is the same for all resonances
in one ladder. In general, the decay rate scales as
$\Gamma \sim F \exp(-\pi \Delta E^2 /F)$, where $\Delta E$ is the
energy gap between the Bloch bands of the periodic potential.
This result can be deduced from Landau-Zener theory \cite{Holt00,00restun}.
However, one observes peaks of the decay rate on top due to
resonant tunneling. For example, the decay rate of the two most
stable resonances for the periodic potential $V(x) = \cos x$
is plotted as a function of the inverse field strength $1/F$
in figure \ref{fig-restunnel0}. RET takes place
when a state of a lower ladder with energy $\E_{\alpha,n}$ gets
in resonance with a state of a higher ladder at a different site,
i.e. $E_{\alpha,n} = E_{\alpha',n'}$. The decay rate of the lower
ladder is significantly increased as it couples resonantly to a
higher ladder with a higher decay rate.
For example, the pronounced peak in the ground ladder decay rate
(i.e. $\alpha=0$) at $F \approx 1/7$ corresponds to the resonance
$\alpha=0 \leftrightarrow \alpha'=1$ and $n'=n-1$.

\subsection{Double-periodic lattices}
\label{sec-nlws-double}

Now we turn to the main subject of the present paper. We consider a
double-periodic potential $V(x)$ consisting of a major optical
lattice of period $d$ plus an additional shallow lattice of double
period,
\be
  V(x) = V_0 \left( \sin^2(\pi x /d) +
  \delta \sin^2(\pi x/2d +\phi/2) \right).
\ee
Rescaling the spatial coordinate as $x' = 2 \pi x/d$ and neglecting a
constant potential offset, we can rewrite the periodic potential as
\be
  V(x) = -\frac{V_0}{2} \left( \cos(x) + \delta \cos(x/2 +\phi) \right).
  \label{eqn-pot-dp}
\ee
The relative phase of the two lattices is denoted by $\phi$.

\begin{figure}[t]
\centering
\includegraphics[width=7cm,  angle=0]{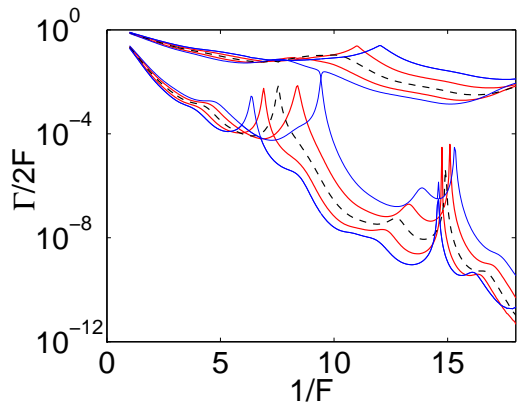}
\caption{\label{fig-restunnel1}
(Color online) Resonant tunneling in a double-periodic Wannier-Stark
system. Decay rates of the four most stable resonance for $V_0 = 2$,
$\phi = 0$ and $\delta = 0$ (dashed black line),
$\delta = 0.05$ (solid red lines) and $\delta = 0.1$
(solid blue lines).}
\end{figure}

Due to the additional lattice each Bloch band splits up into
two minibands \cite{G1998}, and each Wannier-Stark ladder splits up into
two miniladders, as proved in the following. The symmetries
of the Hamiltonian are given by the commutation relations
\begin{eqnarray}
  && [H,T_{2m}] = -2mdF T_{2m}\nn \\
  && [H,T_{2m+1} G] = -(2m+1)dF T_{2m+1},
\end{eqnarray}
where $T_m$ is the translation operator over $m$ lattice periods
and the operator $G$ inverts the sign of $\delta$ in all following
terms. Then it is easy to see that the Wannier-Stark states of one ladder
are related by a translation over an even number of lattice periods,
or by a translation over an odd number of lattice periods plus an
inversion of the sign of $\delta$,
\begin{eqnarray}
  && H T_{2m} \Psi_{\alpha,n}(x) \nn \\
  && \quad  = \left(\E_{\alpha,n}(\delta) - 2m dF\right)  T_{2m} \Psi_{\alpha,n}(x),  \\
  && H T_{2m+1} G \Psi_{\alpha,n}(x) \nn \\
  && \quad = \left(\E_{\alpha,n}(-\delta) - (2m+1) dF\right)  T_{2m+1} G \Psi_{\alpha,n}(x). \nn
\end{eqnarray}
Furthermore, it can be shown that the energy offset $\E_{\alpha,0}$
is antisymmetric in $\delta$,
\be
  \E_{\alpha,0}(-\delta) = - \E_{\alpha,0}(\delta).
\ee
Thus the Wannier-Stark ladders split up into two miniladders, each with an
energy offset $2 \E_{\alpha}(\delta)$:
\begin{eqnarray}
  \E_{\alpha,2n} &=& \E_{\alpha}(\delta) + 2ndF \nn \\
  \E_{\alpha,2n+1} &=& -\E_{\alpha}(\delta) + (2n+1)dF.
\end{eqnarray}
A similar proof is given in \cite{06bloch_zener} within the tight-binding
approximation.

\begin{figure}[t]
\centering
\includegraphics[width=7cm,  angle=0]{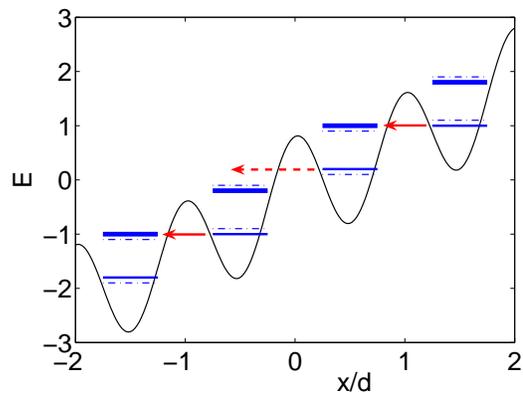}
\caption{\label{fig-res-explain}
(Color online) Explanation of resonantly enhanced tunneling (RET).
See text for details.}
\end{figure}

The decay of the Wannier-Stark resonances is seriously influenced
by the additional period-doubled potential.
Figure \ref{fig-restunnel1} shows the decay rate $\Gamma$ as
a function of the inverse field strength $1/F$ for $V_0 = 2$,
$\phi = 0$ and $\delta = 0.05$ and $\delta = 0.1$, respectively.
The decay rate of the
single-periodic lattice $\delta=0$ is also plotted for comparison.
As all Wannier-Stark ladders split up into two miniladders,
so does the decay rate $\Gamma$. The general scaling of $\Gamma$
with $F$ remains the same for both miniladders, while the RET
peaks are seriously altered. The peaks split up into two,
where the height of the subpeaks increases significantly. This effect
is mostly pronounced for the major resonance at $F \approx 1/7$.
The explanation of the splitting is straightforward: In figure
\ref{fig-res-explain}, the solid blue lines represent the energy
levels of the two most stable Wannier-Stark ladders. Due to the
period-doubled potential they are alternately shifted up or down
with respect to the unperturbed level ($\delta=0$, dash-dotted
lines). The decay rate $\Gamma$ is resonantly enhanced when two
energy levels of different ladders are degenerate, as indicated
by the solid red arrow. However, for $\delta \ne 0$, the two
miniladders are in resonance with higher ladders for different values
of the field strength $F$ due to the alternating energy shift.
In figure \ref{fig-res-explain}, for example, one of the miniladders
is off resonance (indicated by the broken arrow) while the other one
is in resonance (solid arrow). If the field strength is lowered, the
off-resonant ladder will get into resonance at another value of $F$.

\begin{figure}[b]
\centering
\includegraphics[width=8cm,  angle=0]{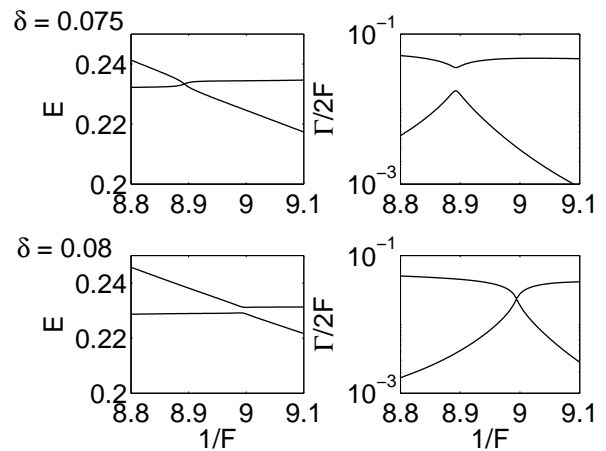}
\caption{\label{fig-wsres-crossing}
(Anti) Crossing of the real and the imaginary part of two eigenenergies
of the Wannier-Stark resonances in a double-periodic Wannier-Stark
system for $V_0 = 2$, $\phi = 0$ and $\delta = 0.075$ (upper panels),
$\delta = 0.08$ (lower panels).}
\end{figure}

With increasing amplitude $\delta$ of the second double-periodic
lattice, the splitting of the resonant tunneling peaks clearly
becomes more pronounced as one can see in figure \ref{fig-restunnel1}.
Another (not so intuitive) effect is that the additional lattice can
also alter the crossing type. For $\delta =0.1$ one observes a
crossing of the decay rates at $F \approx 1/9.5$ due to resonant
tunneling instead of an anti-crossing.
This effect is further illustrated in figure \ref{fig-wsres-crossing}.
The real and the imaginary part of the resonance eigenenergies are plotted
in the vicinity of one of the RET peaks (cf. figure \ref{fig-restunnel1}).
For $\delta = 0.075$, one observes a familiar RET peak,
i.e., a type II curve crossing. For $\delta = 0.08$, however, the crossing
type is altered from type II to type I. The decay rate of one
miniladder crosses the decay rate of one excited miniladder. Correspondingly,
the real parts anti-cross.
A diabolic point, where real and imaginary part are degenerate, is found at
$\delta = 0.0772$ and $F = 1/8.937$ for $\phi = 0$.
If we consider the relative phase $\phi$ as another free parameter,
the set of diabolic points is a one-dimensional subset of the 
three-dimensional parameter space $(\delta,F,\phi)$;
the diabolic crossing has codimension one.

\subsection{Output control by the relative phase}

\begin{figure}[t]
\centering
\includegraphics[width=7cm,  angle=0]{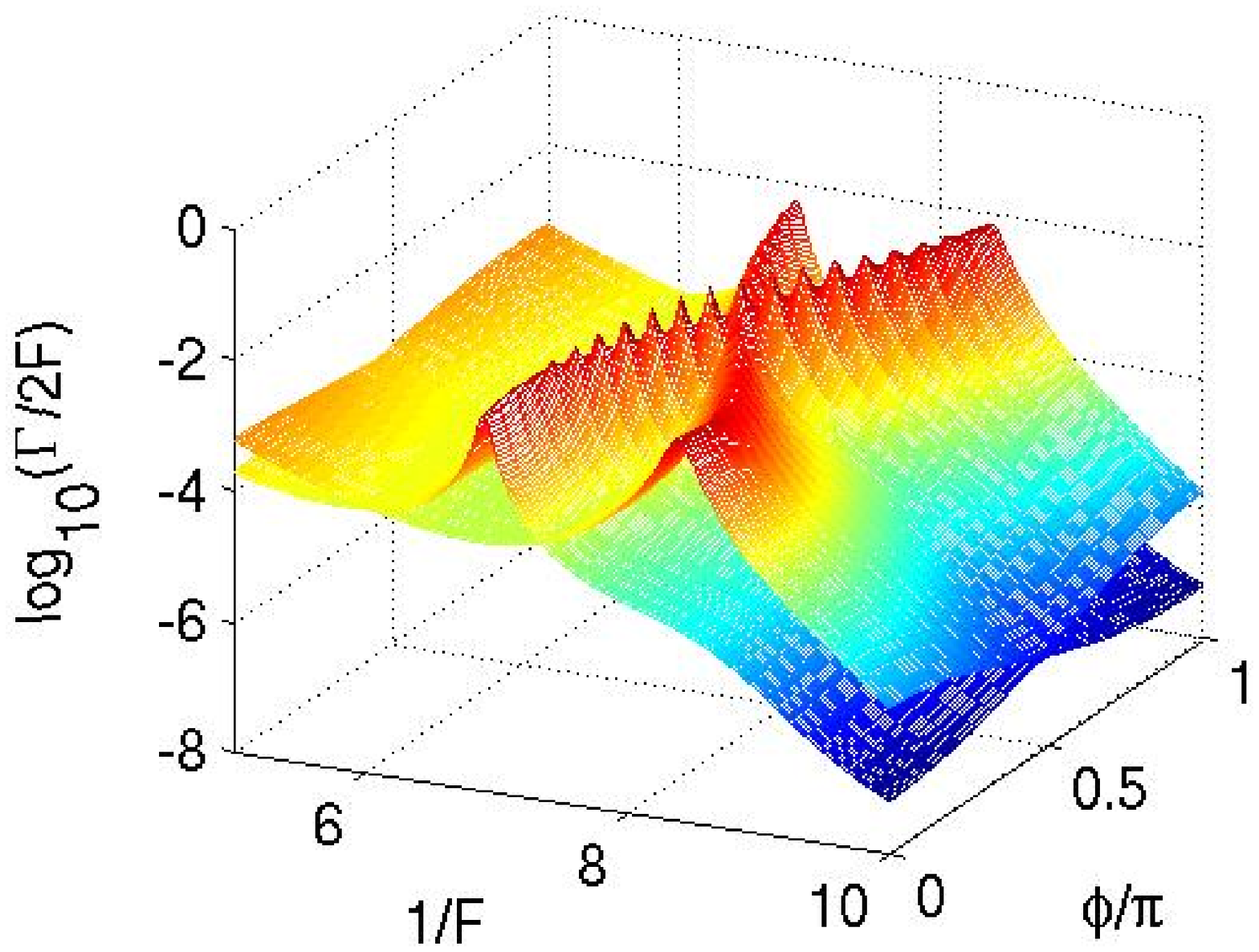}
\includegraphics[width=7cm,  angle=0]{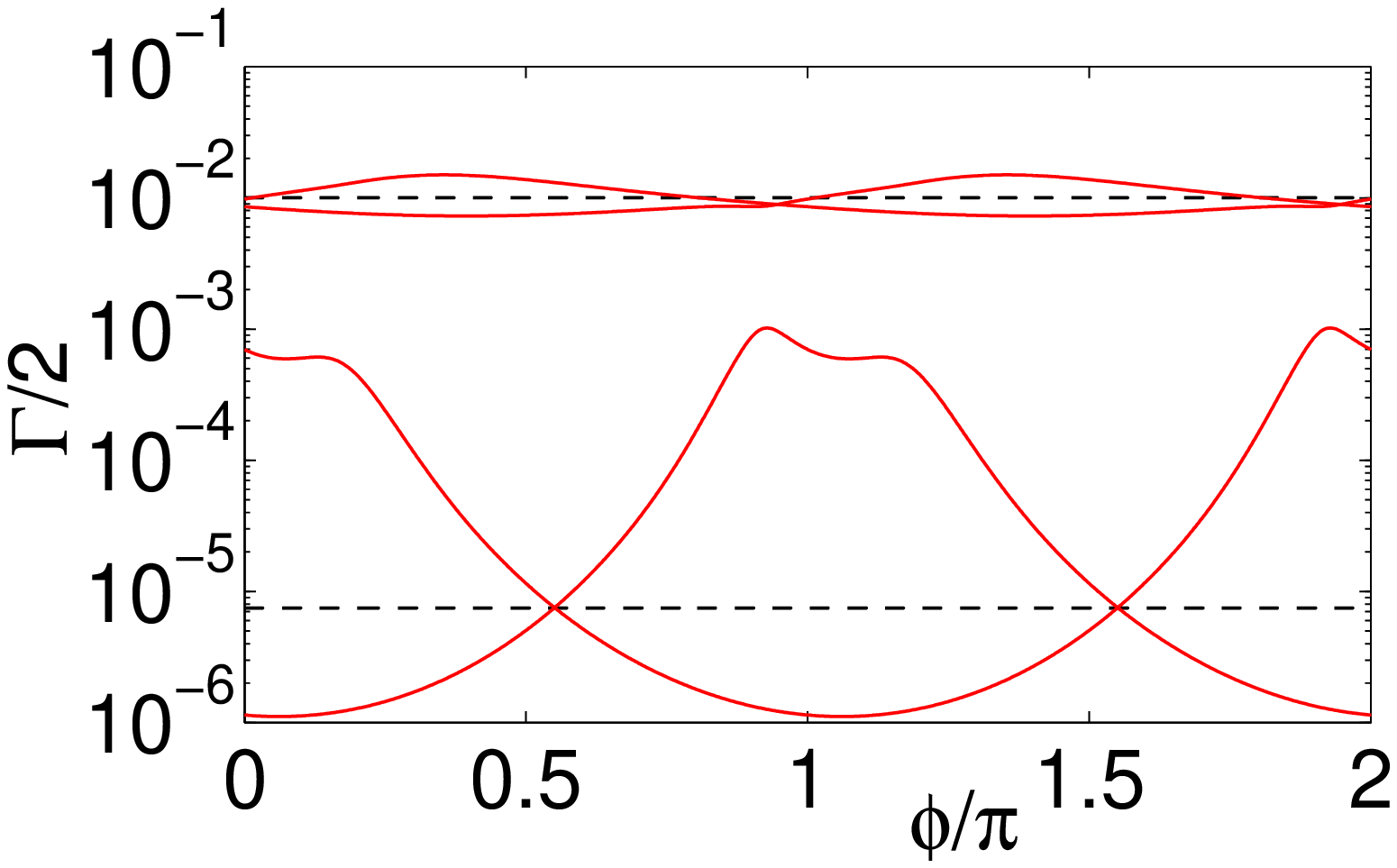}
\caption{\label{fig-res-phase1} \label{fig-res-phase2}
(Color online) Upper panel:
Decay rates of the two most stable resonances as a function of
the relative phase $\phi$ and the inverse field  strength $1/F$
for $V_0 = 2$ and $\delta = 0.05$. Lower panel: For fixed
Stark field $F = 1/9$.}
\end{figure}

Up to now, we have shown that the RET peaks split
up, whereby the splitting increases with the amplitude $\delta$
of the additional lattice. Furthermore, the decay depends crucially
on the relative phase $\phi$ of the lattices.
Figure \ref{fig-res-phase1} shows the decay rate of the two lowest
miniladders in dependence of the relative phase $\phi$ and the
inverse field strength $1/F$ for $V_0 = 2$ and $\delta = 0.05$.
The splitting of the RET peaks is maximal for
$\phi = 0$. It becomes zero for $\phi \approx \pi/2$, where the
decay rates of the two miniladders degenerate again.
Despite the fact that the additional lattice is much weaker than the
single-periodic one ($\delta = 0.05$), it seriously affects the
decay properties. The decay rate of the two lowest miniladders vary
over several orders of magnitude.
This is further illustrated in the lower panel of figure
\ref{fig-res-phase1}, where the decay rate is plotted for
a fixed value of $F=1/9$.
The decay rate for the single-periodic case $\delta=0$ is also plotted
for comparison. The decay rate as a function of the phase,
$\Gamma(\phi)$, is $\pi$-periodic. This can be seen as follows:
Shifting the phase by an amount of $-\pi$ and the position by $2\pi$
leaves the system invariant. A spatial translation by $2 \pi$
exchanges the two miniladders.

\begin{figure}[t]
\centering
\includegraphics[width=8cm,  angle=0]{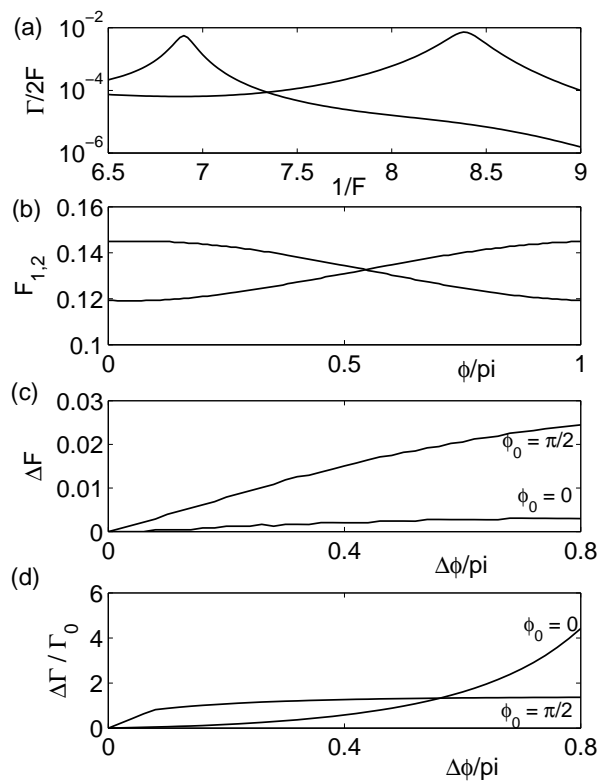}
\caption{\label{fig-res-ws-jitter1}
Effects of fluctuations of the relative phase $\phi$.
(a) Decay rate of the two most stable resonances for $V_0 = 2$,
$\delta = 0$ and $\phi=0$. (b) Position $F_{1,2}$ of the RET
peaks vs. the relative phase $\phi$.
(c) Uncertainty $\Delta F$ of the positions of the RET
peaks vs. fluctuation $\Delta \phi$ of the phase.
(d) Relative fluctuation $\Delta \Gamma / \Gamma_0$ of the decay
rate for $F=1/8$ vs. $\Delta \phi$.}
\end{figure}

The sensitive dependence on the relative phase $\phi$ might be
very useful for the design and control of future experiments with
double-periodic optical lattices. A complete stabilization of the
relative phase of the two lattices will be hard to achieve and fluctuations
will play an important role.
Yet, Ritt~et~al. have realized a new technique based on Fourier
synthesis to control two (or even more) optical lattices with different
spatial harmonics \cite{Ritt06}.
In the following, we will discuss the effects on the decay rates of
the Wannier-Stark resonance in some detail. The strongest effects
are found in the vicinity of the RET peaks, on which we focus in
our present discussion.

Figure \ref{fig-res-ws-jitter1} (a) again presents the decay rate as a
function of the inverse field strength $1/F$ for $V_0 = 2$ and
a weak additional lattice with $\delta = 0.05$ and $\phi = 0$.
For the given parameters, we find RET peaks at
$F_1 \approx 1/6.9$ and $F_2 \approx 1/8.4$.
Figure \ref{fig-res-ws-jitter1} (b) shows the position $F_{1,2}$
of the two the resonant tunneling peaks in dependence of the
relative phase $\phi$. The positions vary in an interval of width
$\Delta F \approx 1/40$.

In a real-life experiment, it is difficult to exactly control the relative
phase of two independent standing waves. Therefore, we study also the
influence of random phase fluctuations.
If we assume that the phase $\phi$ fluctuates in an interval
of width $\Delta \phi$ around the desired value $\phi_0$,
$\phi \in [\phi_0 - \Delta \phi/2,\phi_0 + \Delta \phi/2 ]$,
the positions of the resonant tunneling peaks will also
fluctuate in an interval of width $\Delta F$.
Figure \ref{fig-res-ws-jitter1} (c) shows how the width $\Delta F$ depends
on the strength of the phase fluctuations $\Delta \phi$
for $\phi_0 = 0$ and $\phi_0 = \pi/2$. The fluctuations are rather
weak for $\phi_0 = 0$, where the RET peaks have maximum distance.
For a given value of the external field $F$, the decay rates $\Gamma$
fluctuate in an interval of width $\Delta \Gamma$.
Figure \ref{fig-res-ws-jitter1} (d) shows the relative strength of the
fluctuation, $\Delta \Gamma / \Gamma_0$ vs. the phase fluctuations
$\Delta \phi$ for the most stable resonance for an external field $F = 1/8$
and $\phi_0 = 0$ or $\phi_0 = \pi/2$, respectively. The relative uncertainty
of the decay rate becomes greater than unity already for small
fluctuations of the phase $\phi$. Nevertheless, the noise induced shift
and change in height of the RET peaks is small considering the
absolute change of the decay rates around the RET peaks of about two orders
of magnitude.

However, it is also possible to exploit the sensitive dependence on
the phase $\phi$, if it can be accurately controlled. For instance, it
could be possible to rapidly tune the output of a pulsed atom laser.
An example will be discussed in detail in section \ref{sec-appl}.

\section{Nonlinear Wannier-Stark resonances in double-periodic optical lattices}
\label{sec-ws-nonl}

\begin{figure}[t]
\centering
\includegraphics[width=7cm,  angle=0]{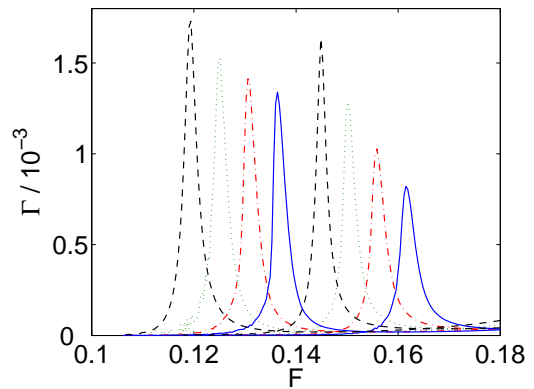}
\caption{\label{fig-nlwsres}
(Color online) Decay rates of nonlinear Wannier-Stark resonances
in double periodic optical lattices with $V_0 = 2$, $\delta = 0.05$,
$\phi = 0$ and $g=0$ (dashed black line), $g=-0.1$ (dashed green)
$g=-0.2$ (dash-dotted red) and $g=-0.3$ (solid blue), respectively.}
\end{figure}

A method to obtain accurate, nonlinear Wannier-Stark resonances was
proposed recently in \cite{Wimb06}, and we use a similar approach to
numerically compute decay rates of the nonlinear version of the
Wannier-Stark introduced in section \ref{sec-nlws-double} above.
In contrast to the case studied in \cite{Wimb06}, the computations
based on the Gross-Pitaevskii equation (\ref{eqn-NLSE-timedep})
in the presence of a two-period optical lattice are more difficult,
since the algorithm needs to discriminate between the two miniband
solutions which are quite close in energy (cf. Figs. \ref{fig-restunnel0}
and \ref{fig-restunnel1}). In particular, for very small Stark
fields, it is hard to obtain convergence. In the following, we
concentrate therefore on RET peaks at fields as large as possible, and one
trick to shift the peaks to such values is to use {\em attractive}
interactions, i.e. negative nonlinearities ($g<0$). Figure \ref{fig-nlwsres}
presents a set of RET peaks for different values of the nonlinearity $g$.

\begin{figure}[b]
\centering
\includegraphics[width=7cm,  angle=0]{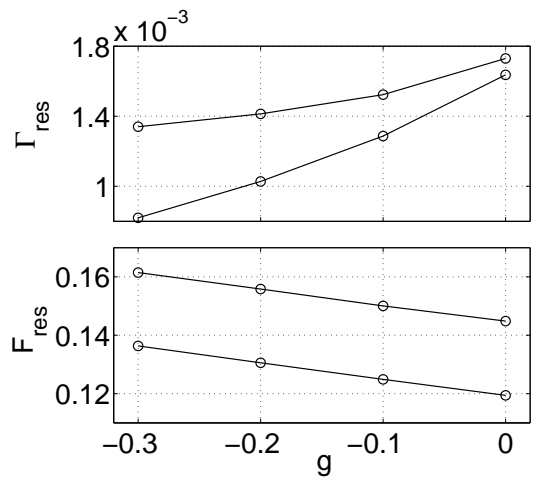}
\caption{\label{fig-peak-fres}
Shift of the RET peaks. Shown is the position $F_{\rm res}$
of the RET peaks (lower panel) and the resonant decay rate $\Gamma_{\rm res}$
(upper panel) in dependence of the nonlinearity $g$.}
\end{figure}

The nonlinearity induced a shift of both RET peaks (corresponding to the two
minibands) and also a systematic stabilization (i.e., smaller heights) can be
observed, as predicted by similar results for the usual, one-band Wannier-Stark
system \cite{Wimb06}.
This is analyzed in more detail in figure \ref{fig-peak-fres}, where the
peak positions $F_{\rm res}$ and the height of the peaks $\Gamma_{\rm res}$
are plotted in dependence of $g$.

The stabilization of the Wannier-stark states by an attractive
nonlinearity is shown in the upper panel.
An asymmetry of the two peaks is observed already in the linear case
$g=0$: The left peak is slightly higher than the right one, i.e. the peak
decay rate is larger for {\it smaller} external fields $F$.
This phenomenon becomes even more pronounced in the nonlinear case $g<0$.
The stabilization by an attractive nonlinearity (cf. \cite{Wimb06})
is stronger for the right peak.

The peak positions shown in the lower panel of figure \ref{fig-peak-fres}
vary linearly with $g$, which can be derived in a perturbative approach.
As discussed above, resonant tunneling is observed when a state of a lower
ladder get in resonance with a state in a higher ladder at a different site,
$E_{\alpha,n} = E_{\alpha',n'}$.
Here we consider only the states in the ground ladder, which are localized
in a single potential well. First order perturbation theory with respect to
the linear case $g=0$ predicts that their energy is shifted
by the amount \cite{Wimb06}
\be
  \Delta E_{0,0} \approx g \int_{-\pi}^{+\pi} |\psi_{0,0}^{(0)}(x)|^4 \rd x,
  \label{eqn-wsnonlin-pert1}
\ee
where the superscript $(0)$ refers to the linear case $g=0$.
The shift $\Delta F$ of the RET peaks then follows from the modified resonance
condition
\be
  E_{0,n}^{(0)} + \Delta E_{0,n} = E_{\alpha',n'}^{(0)} + n' d \, \Delta F.
\ee
Evaluating the integral in equation (\ref{eqn-wsnonlin-pert1}) and setting
$n' = n-1$ one finds
\be
  \Delta F = - \frac{0.36}{2 \pi} g
\ee
for the peak shift plotted in figure \ref{fig-peak-fres}. Both peaks are
shifted equally, so that the distance of the peaks remains constant.

\begin{figure}[t]
\centering
\includegraphics[width=7cm,  angle=0]{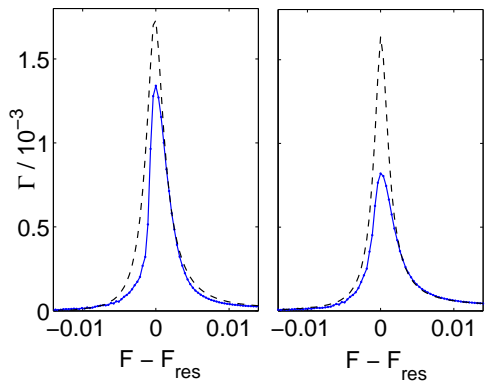}
\caption{\label{fig-peakform1}
(Color online) Asymmetry of the nonlinear RET peaks.
Shown is the decay rate $\Gamma(F)$ of the two peaks (left and right)
around the respective peak positions $F_{\rm res}$ for $g=0$ (black
dashed) and $g=-0.3$ (solid blue line).}
\end{figure}

Furthermore the shape of the RET peaks becomes asymmetric in the nonlinear
case. This is shown in figure \ref{fig-peakform1}, where we have plotted
a magnification of the decay rate $\Gamma(F)$ around the respective
positions of the RET peaks for $g=-0.3$ and $g=0$. In comparison to the
linear case, the peak is bent to the left for an attractive nonlinearity.
This asymmetry is a general feature of nonlinear eigenstates in open systems.
It is already present for the nonlinear two-level system as shown in
figure \ref{fig-2level-nlin-cross12weakc}.
A similar incline of resonant curves is also important for nonlinear resonant
transport. The curves can even bend over for strong nonlinearities leading
to a bistable behavior as shown in \cite{Paul05nlt}.

\section{Dynamics}
\label{sec-appl}

In this section, we discuss the dynamics of an initially localized matter
wave, e.g., a Gaussian wave packet in the tilted double-periodic
optical lattice (\ref{eqn-pot-dp}).
In one of the first experiments on the macroscopic dynamics of BECs
in optical lattices it was shown that such a system shows a coherent
pulsed output \cite{Ande98}.
An explanation in terms of truncated Wannier-Stark resonances
can be found in \cite{02pulse}. The amplitude of the pulsed output
is given by the decay rate of the Wannier-Stark resonances.

\begin{figure}[t]
\centering
\includegraphics[width=7cm,  angle=0]{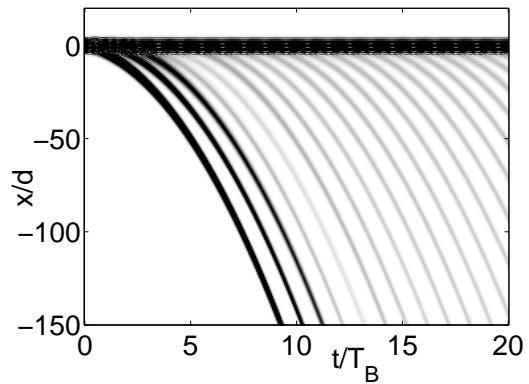}
\caption{\label{fig-wspulse1}
Pulsed output from a tilted optical lattice for $V_0 = 0.8$, $F=1/18$ and
$\delta = 0, g=0$. Shown is the atomic density $|\psi(x,t)|^2$ in a
greyscale plot.}
\end{figure}

\begin{figure}[t]
\centering
\includegraphics[width=7cm,  angle=0]{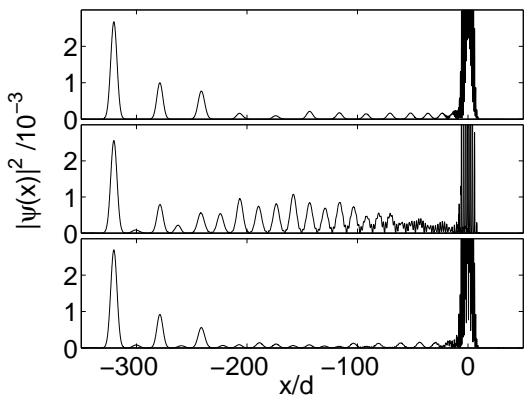}
\caption{\label{fig-wspulse2}
Pulsed output from a tilted optical lattice for $g=0$. Shown is the atomic
density for $t = 14 \, T_B$ for $\delta = 0$ (top), $\delta = 0.2$
and $\phi = 0$ (middle) and $\delta = 0.2$ and $\phi = \pi/2$ (bottom).}
\end{figure}

First of all, we illustrate how the sensitive dependence on the phase
can be be used to tune a pulsed atom laser. As a proof of principle,
we just consider the {\it linear} ($g=0$) evolution.
In contrast to the previous sections we consider a weaker
potential, $V_0 = 0.8$ so that decay is generally stronger.
We numerically integrate the Schr\"odinger equation for an initially
Gaussian wave packet
\be
  \psi(x,t=0) = \frac{1}{(2\pi)^{1/4} \, \sigma^{1/2}}  \;
  \exp(-(x-x_0)^2/4\sigma^2)
\ee
with width $\sigma = 5\pi$. Figure \ref{fig-wspulse1} shows the density
$|\psi(x,t)|^2$ in a greyscale plot for a single periodic lattice
($\delta = 0$). A pulsed output forms due to the
external field. The pulses are accelerated just as
classical particles. The first three strong pulses emerge from excited
ladders. The output strength of the other pulses from the ground ladder
can be controlled to a large extent in the double-periodic case by the
relative phase $\phi$.
As already shown in figure \ref{fig-res-phase1}, the decay rate varies
strongly with the relative phase of the two lattices. The effect on
the pulsed output is shown in figure \ref{fig-wspulse2}, where the
density is plotted for $t= 14 T_B$ for three different lattice setups.

For $\delta = 0.2$ and $\phi = 0$, a RET peak of type I is found in the 
second miniladder at $F = 1/18$. Thus decay from this miniladder is strongly
enhanced in comparison to the single-periodic lattice $\delta = 0$.
Note that the pulsed output will stop as soon as the population in the 
second miniladder has decayed.
In contrast, the pulsed output is strongly suppressed for a relative phase
$\phi = \pi/2$, where RET does not play a role for the given field strength.

In order to measure the performance of this output switch more qualitatively,
we define the fidelity
\be
  f = \frac{P_{\rm out}(\phi_0 = 0)}{P_{\rm out}(\phi_0 = \pi/2)},
\ee
where $P_{\rm out}$ measures the integrated density of the pulsed output
for a certain value $\phi_0$ of the relative phase of the two
optical lattices. The output is switched on for $\phi_0 = 0$ and
it is switched off for $\phi_0 = \pi/2$.
We measure the output density $P_{\rm out}$ at $t= 14 T_B$, where we
neglect the first three strong pulses as they are due to the initial
population of excited Bloch bands. Then one has
\be
  P = \int_{-230d}^{-30d} |\psi(x)|^2 \rd x.
\ee
For the parameters used in figure \ref{fig-wspulse2}, we find a fidelity
of $F = 7.4$, i.e., the output for $\phi_0 = 0$ is enhanced by a factor
of $7.4$ in comparison to $\phi_0 = \pi/2$.

\begin{figure}[t]
\centering
\includegraphics[width=7cm,  angle=0]{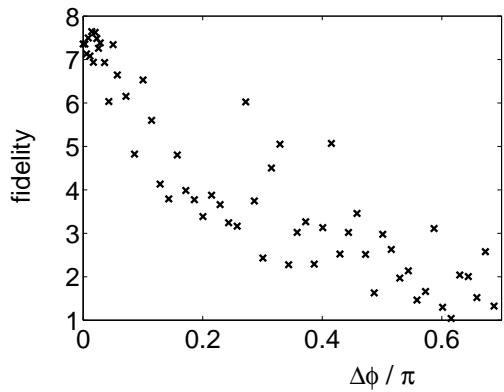}
\caption{\label{fig-wspulse3}
Fidelity of the output switch in dependence of the standard deviation
$\Delta \phi$ of the phase fluctuations.}
\end{figure}

However, in a real experimental setup the phase $\phi$ will fluctuate
around the desired value.
This is mainly due to mechanical perturbations, thus fluctuations
with very high frequencies are unlikely, while fluctuations with small
frequencies up to some kHz can be controlled by an active stabilization.
Thus we assume that the power spectrum of the phase fluctuations has a
maximum at intermediate values in the kHz regime. This is comparable to
the Bloch frequency $\omega_B = 2\pi/T_B$, as the Bloch period is about
one millisecond in a typical experiment \cite{Ande98}.
Exemplarly, we consider fluctuations with a Gaussian power spectrum
with mean $\omega_B$ and width $\omega_B/4$.
In the following we analyze the pulsed output in
dependence of the strength of the fluctuations. Figure \ref{fig-wspulse3}
shows the fidelity of the output in dependence of the standard deviation
$\Delta \phi = (\langle (\phi - \phi_0)^2 \rangle)^{1/2}$
of the fluctuations for the same parameters as in figure \ref{fig-wspulse2}.
One observes that the fidelity drops to one (no switching effect) for
a standard deviation of $\Delta \phi \approx 0.6 \pi$. A reduction of
the fluctuations below this value is in principle possible today, however
only with a great technical effort. As a consequence, an output switching
seems feasible in double periodic lattices.

\section{Conclusion and Outlook}

In the present paper we have studied the interplay between decay and
a nonlinear mean-field potential describing the atom-atom
interactions in a dilute Bose-Einstein condensate.

As an illustrative model we have investigated a two-level systems
with one decaying level, which can be treated analytically.
In the linear case, one has to distinguish two types of level
crossings, either the real parts anticross while the imaginary parts
of the eigenvalues cross (type I) or the other way around (type II).
Both real and imaginary parts are degenerate at the exceptional point,
where the bare state decay rate equals the coupling strength.
A weak nonlinearity does {\it not} alter the crossing type, however it
{\it deforms} the levels in a characteristic manner. For example,
it leads to a bending of the peaks in the decay rates.
Novel nonlinear eigenstates emerge for a stronger nonlinearity,
where the critical nonlinearity is decreased in the presence of
decay. Looped levels appear for type I crossings, while the additional
eigenstates emerge in a double S-structure for a type II crossing.
At the exceptional point novel eigenstates emerge, even for small
nonlinearities.

An experimental setup where both decay and nonlinearity
play an important role is the dynamics of Bose-Einstein condensates
in accelerated optical lattices. In particular, we have analysed the
decay in a double-periodic lattice, where a weak period-doubled
potential is superimposed onto the fundamental lattice. These results
will be of interest for controlling transport of ultracold atoms in
future and on-going experiments \cite{Pisa06}.

The decay rate in a double-periodic Wannier-Stark system depends sensitively
on the system parameters, such as the relative amplitudes of the lattices and
the relative phase, which can be varied over a wide range.
In particular, the resonant tunneling peaks of the decay rate
$\Gamma(F)$ split up into two subpeaks. Varying the system parameters
one can tune these peaks and even achieve a crossover from a
type II crossing to a type I crossing.
This could be crucial for future experiments since the a robust
control of the relative phase is hard to realize. One can,
however, also exploit this sensitive dependence in order to
implement a fast output switch for a pulsed atom laser.
A weak nonlinear mean-field potential describing the atom-atom
interactions in a Bose-Einstein condensate of ultracold atoms
has two major effects: The resonant tunneling peaks are shifted.
This shift can lead to a stabilization against decay. Furthermore
it leads to a bending of the peaks as predicted by the
nonlinear non-hermitian two-level system.


\begin{acknowledgments}
We thank Peter Schlagheck for fruitful and inspiring discussions.
DW acknowledges support from the Studienstiftung des deutschen Volkes
and the Deutsche Forschungsgemeinschaft via the Graduiertenkolleg 792.
SW acknowledges support from the Alexander von Humboldt Foundation
(Feodor-Lynen Program).

\end{acknowledgments}


\end{document}